\documentclass[a4paper,10pt]{article}

\usepackage{epsf}
\usepackage{graphicx}

\newcommand{\be}[1]{\begin{equation}\label{#1}}
\newcommand{\ee}{\end{equation}}
\newcommand{\bea}{\begin{eqnarray}}
\newcommand{\eea}{\end{eqnarray}}

\def\gsim{ \lower .75ex \hbox{$\sim$} \llap{\raise .27ex \hbox{$>$}} }
\def\lsim{ \lower .75ex \hbox{$\sim$} \llap{\raise .27ex \hbox{$<$}} }

\renewcommand{\markright}{\markright{\thepage}}

\begin{document}

\begin{titlepage}

\begin{center}
{\qquad \qquad \qquad \qquad \qquad \qquad \qquad \qquad \qquad
\qquad \qquad \qquad \qquad \qquad \qquad \qquad \qquad
CAS-KITPC/ITP-016}
\end{center}

\vspace{5mm}

\begin{center}

{\Large \bf On Perturbations of a Quintom Bounce}

\vspace{10mm}

{\large Yi-Fu Cai$^1$\footnote{caiyf@mail.ihep.ac.cn}, Taotao
Qiu$^1$\footnote{qiutt@mail.ihep.ac.cn}, Robert
Brandenberger$^{2,3}$\footnote{rhb@hep.physics.mcgill.ca},
Yun-Song Piao$^4$\footnote{yspiao@gucas.ac.cn}, Xinmin
Zhang$^{1,5}$\footnote{xmzhang@mail.ihep.ac.cn}}

\vspace{5mm}

{\em \small
$^{1}$ Institute of High Energy Physics, Chinese Academy of Sciences, P.O. Box 918-4, Beijing 100049, P. R. China\\
$^{2}$ Department of Physics, McGill University, Montr¡äeal, QC, H3A 2T8, Canada\\
$^{3}$ Kavli Institute for Theoretical Physics, Zhong Guan Cun East Street
55, Beijing 100080, P.R. China\\
$^{4}$ College of Physical Sciences, Graduate School of Chinese
Academy of Sciences, Beijing 100049, P. R.
China\\
$^{5}$ Theoretical Physics Center for Science Facilities (TPCSF),
CAS, P. R. China}

\end{center}

\vspace{5mm}

\begin{abstract}
A Quintom universe with an equation-of-state crossing the
cosmological constant boundary can provide a bouncing solution
dubbed the Quintom Bounce and thus resolve the Big Bang
singularity. In this paper, we investigate the cosmological
perturbations of the Quintom Bounce both analytically and
numerically. We find that the fluctuations in the dominant mode in
the post-bounce expanding phase couple to the growing mode of the
perturbations in the pre-bounce contracting phase.
\end{abstract}

\end{titlepage}

\newpage

\setcounter{page}{2}

\section{Introduction}

For years, it has been suggested that bouncing universe scenarios
could provide a solution to the problem of the initial singularity
in the evolution of the cosmos, a problem which afflicts both
Standard Big Bang cosmology and inflationary universe models
\cite{Guth81,Sato,Steinhardt82,Linde82} \footnote{An initial
singularity occurs in inflationary models if scalar fields are
used to generate  the accelerated expansion of space \cite{Borde}.
If inflation arises as a consequence of modifications of the
gravitational part of the action, as in Starobinsky's original
proposal \cite{Starob}, it is possible that the evolution is
singularity-free.}. In the literature there have been a lot of
efforts towards constructing bouncing universes, for instance the
Pre-Big-Bang scenario \cite{PBB} or the cyclic/Ekpyrotic scenario
\cite{Ekp}. Also, in loop quantum cosmology the evolution might be
singularity-free \cite{Bojowald:2001xe}. In Refs.
\cite{Brustein:1997cv,Cartier,Tsujikawa} and
\cite{Biswas:2005qr,Setare:2004jx} models were considered in which
the gravitational action is modified by adding high order terms.
Since, in general, in these models the equations for cosmological
perturbations are very involved, it is difficult to make
definitive conclusions about the evolution of fluctuations, and
the knowledge of the evolution is crucial in order to compare the
predictions of the model with observations.

In some previous work \cite{Cai:2007qw}, a subset of the present
authors considered a bouncing universe model (``Quintom Bounce")
obtained within the standard 4-dimensional
Friedmann-Robertson-Walker (FRW) framework, but making use of
Quintom matter \cite{Feng:2004ad}. In this model, it can be shown
in \cite{Cai:2007qw} that the null energy condition (NEC) is
violated for a short period around the bounce point. Moreover,
after the bounce the equation-of-state (EoS) of the matter content
$w$ in our model is able to transit from $w<-1$ to $w>-1$ and then
connect with normal expanding history.

A Quintom model \cite{Feng:2004ad} was initially proposed to
obtain a model of dark energy with an EoS parameter $w$ which
satisfies $w>-1$ in the past and $w<-1$ at present. This model is
mildly favored by the current observational data fitting
\cite{Zhao:2006qg}. The Quintom model is a dynamical scenario of
dark energy with a salient feature that its EoS can smoothly cross
over the cosmological constant barrier $w = -1$. It is not easy to
construct a consistent Quintom model. A no-go theorem proven in
Ref. \cite{Xia:2007km} (also see Refs.
\cite{Feng:2004ad,Vikman:2004dc,Hu:2004kh,Caldwell:2005ai,Zhao:2005vj,Abramo:2005be,Kunz:2006wc})
forbids a traditional scalar field model with Lagrangian of the
general form ${\cal L} = {\cal
L}(\phi,\partial_{\mu}\phi\partial^{\mu}\phi)$ from having its EoS
cross the cosmological constant boundary. Therefore, it is
necessary to add extra degrees of freedom with un-conventional
features to the conventional single field theory if we expect to
realize a viable Quintom model in the framework of Einstein's
gravity theory. The simplest Quintom model involves two scalars
with one being Quintessence-like and another Phantom-like
\cite{Feng:2004ad,Guo:2004fq}. This model has been studied in
detail in \cite{Quintom_tf}. In recent years there have been a lot
of theoretical studies of Quintom models. For example, motivated
from string theory, the authors of Ref. \cite{Cai:2007gs} realized
a Quintom scenario by considering the non-perturbative effects of
a DBI model. Moreover, there are models which involve a single
scalar with higher derivative terms in the action
\cite{Li:2005fm}, models with vector field
\cite{ArmendarizPicon:2004pm}, making use of an extended theory of
gravity \cite{Cai:2005ie}, of ideas from non-local string field
theory theory \cite{Aref'eva:2005fu}, and others (see e.g.
\cite{Quintom_1,Onemli:2004mb,Quintom_others}). Because of its
brand new properties, Quintom models give many unexpected
predictions. Most important for the subject of the present paper
is that a universe dominated by Quintom matter can provide a
bouncing cosmology which allows us to avoid the problem of the
initial singularity. This scenario, developed in
\cite{Cai:2007qw}, is called  the Quintom Bounce. In this
scenario, the EoS crosses the cosmological constant boundary twice
around the bounce point.

In order to be able to compare the predictions of a Quintom bounce
with observations, it is crucial to investigate the evolution of
cosmological perturbations. In singular bouncing models the
evolution of fluctuations is not under control classically. The
fluctuations diverge at the bounce point, firstly invalidating the
applicability of the theory of linear cosmological perturbations,
and secondly making it impossible to reliably connect the
fluctuations in the contracting phase with those in the expanding
phase. In a Quintom bounce model the evolution of fluctuations is
well-behaved. Our first main result is that cosmological
perturbations can evolve smoothly from the contracting to the
expanding phase. We show that the commonly used variable $\Phi$,
the metric perturbation in longitudinal gauge, passes smoothly
both through the bounce point and through the points where the
cosmological constant boundary of the EoS is crossed.
Interestingly, we find that our result for the transfer of
fluctuations through the bounce possesses features found both in
some previous analyses of fluctuations in non-singular
cosmologies, and features of the evolution in singular bouncing
models. For small comoving wave numbers one finds an evolution
similar to what occurs in singular bouncing cosmologies (the
growing mode of $\Phi$ in the contracting phase couples only to
the decaying mode in the expanding phase), whereas the transfer of
the perturbations is very different in the region of large
comoving wave numbers (the growing mode in the contracting phase
couples to the dominant mode in the phase of expansion). The
physical reason is easy to see: small $k$ modes enter the
sub-Hubble region very close to the bounce point and have no time
to complete a single oscillation, whereas large $k$ modes undergo
many oscillations which are in the bounce phase at sub-Hubble
scales. Thus, whereas the large $k$ modes feel the resolution of
the singularity, the small $k$ modes evolve as if there still were
an abrupt transition between the contracting to the expanding
phase. We will follow the fluctuations both numerically and
analytically. In the analytical analysis, we apply matching
conditions away from the bounce point, in regions where both the
background and the perturbations can be matched consistently via
the Deruelle-Mukhanov \cite{Deruelle} (initially derived in
\cite{Hwang}) conditions \footnote{For a discussion of the dangers
in applying the matching conditions at the transition point
between contraction and expansion see \cite{Durrer}.}.

This paper is organized as follows. In Section 2, we review the
equations of motion of double-field Quintom and derive the
perturbation equations for this model. In Section 3, we study the
behavior of a specific Quintom Bounce model systematically. In
Subsection 3.1, we determine the complete evolution of the
background using numerical integration. In Subsection 3.2, we
investigate the perturbations, first analytically by solving the
equations in separate periods by means of approximations valid
during those periods, and matching the solutions at the transition
points, and finally numerically. The numerical results support our
analytical calculations consistently for all times. Section 4
contains discussion and conclusions.

\section{Review of the Quintom Bounce in a Double-Field Model}

\subsection{Equations of Motion of Double-Field Quintom}

To start, we take a Quintom model consisting of two fields with
the Lagrangian
\begin{eqnarray}
{\cal L} \, = \,
\frac{1}{2}\partial_{\mu}\phi\partial^{\mu}\phi-\frac{1}{2}\partial_{\mu}\psi\partial^{\mu}\psi-V(\phi)-W(\psi)~,
\end{eqnarray}
where the signature of the metric is $(+,-,-,-)$. Here the field
$\phi$ has a canonical kinetic term, but $\psi$ has a kinetic term
with the opposite sign and thus plays a role of a ghost field. In
the framework of a flat Friedmann-Robertson-Walker (FRW) universe,
the metric is given by $ds^2=dt^2-a^2(t) dx^i dx^i$. By varying
the corresponding matter action, we easily obtain the following
expressions for the energy density $\rho$ and pressure $p$ of this
model,
\begin{eqnarray}
\rho=\frac{1}{2}{\dot\phi}^2-\frac{1}{2}{\dot\psi}^2+V(\phi)+W(\psi)~,~~~~
p=\frac{1}{2}{\dot\phi}^2-\frac{1}{2}{\dot\psi}^2-V(\phi)-W(\psi)~,
\end{eqnarray}
and the background equations of motion are given by
\begin{eqnarray}
&H^2=\frac{8\pi
G}{3}[\frac{1}{2}\dot\phi^2-\frac{1}{2}\dot\psi^2+V+W]~,&\\
\label{doyHbg}&\dot H=-4\pi G(\dot\phi^2-\dot\psi^2)~,&\\
\label{phibg} &\ddot\phi+3H\dot\phi+V_{,\phi}=0~,&\\
&\ddot\psi+3H\dot\psi-W_{,\psi}=0~,&
\end{eqnarray}
where $a(t)$ is the scale factor, $H=\frac{\dot a}{a}$ is the
Hubble parameter and the dot denotes the derivative with respect
to the cosmic time $t$.

It is well known that this model can realize the EoS $w$ of a
universe which crosses the cosmological constant boundary. Namely,
the expression for $w$ is given by
\begin{eqnarray}
w=-1+\frac{\dot\phi^2-\dot\psi^2}{\frac{1}{2}{\dot\phi}^2-\frac{1}{2}{\dot\psi}^2+V(\phi)+W(\psi)}~.
\end{eqnarray}
If $\dot\phi^2-\dot\psi^2$ is positive, $w$ is larger than $-1$
and the universe lies in a Quintessence-like phase; if
$\dot\phi^2-\dot\psi^2$ is negative, $w$ is less than $-1$ and
correspondingly the universe is Phantom-like. In recent work, it
was found that the Quintom model can provide a bouncing universe
scenario\cite{Cai:2007qw}. If we work in the context of a
spatially flat four-dimensional background FRW metric, there must
be a period when the Null Energy Condition (NEC) is violated if we
are to obtain a smooth transition from a contracting universe into
an expanding phase. In this case, we need a kind of matter which
admits an EoS parameter which is less than $-1$, but only around
the bounce. Neither  regular not Phantom matter alone can achieve
a transition in the EoS parameter through the cosmological
constant boundary. Therefore, a Quintom model is the only possible
solution to resolve this difficulty.

\subsection{Equations of Motion of Perturbations}

Now let us consider linear perturbations of the metric. The
longitudinal (conformal-Newtonian) gauge metric perturbations are
given by
\begin{eqnarray}
ds^2=a^2(\eta)[(1+2\Phi)d\eta^2-(1-2\Psi)dx^i dx^i]~,
\end{eqnarray}
where we introduced the comoving time $\eta$ defined by
$d\eta=dt/a$ for convenience. The fields $\Phi$ and $\Psi$ depend
on space and time and contain the information about the scalar
metric fluctuations, the degrees of freedom in the metric which
couple to matter fluctuations (see e.g. \cite{MFB} for a
comprehensive survey of the theory of cosmological perturbations
and \cite{RHBrev} for a recent overview). We will not discuss
vector fluctuations since they are not induced by scalar field
matter. To linear order, tensor metric fluctuations (gravitational
waves) do not couple to matter and hence we will not discuss them
here. By expanding the Einstein equations to first order, we
obtain the following perturbation equations
\begin{eqnarray}
\label{peqom1}\nabla^2\Psi-3{\cal H}(\Psi'+{\cal H}\Phi)&=&4\pi G a^2 \delta\rho~,\\
\label{peqom2}\Psi'+{\cal H}\Phi&=&-4\pi G a \delta q~,\\
\label{peqom3}\Phi''+3{\cal H}\Phi'+(2{\cal H}'+{\cal H}^2)\Phi&=&
4\pi G a^2 \delta p~,
\end{eqnarray}
where ${\cal H}=da/ad\eta$ is the comoving Hubble parameter and
the prime denotes the derivative with respect to the comoving
time. Here $\delta\rho$, $\delta q$ and $\delta p$ represent the
perturbations of density, momentum and pressure, respectively. The
first equation is the perturbed `$00$' equation, the second the
perturbed `$0i$' equation, and the third is the diagonal `$ii$'
equation. For scalar field matter there is no anisotropic stress
to linear order in the matter fluctuations, and hence it follows
from the off-diagonal `$ij$' perturbed Einstein equations that
$\Phi=\Psi$.

From the conservation of the stress tensor, we also obtain the
analog of the continuity equation for the matter fluctuations:
\begin{eqnarray}
\delta\rho'+3{\cal H}(\delta\rho+\delta
p)-\frac{1}{a}\nabla^2\delta q=3(\rho+p)\Phi'~.
\end{eqnarray}

In the double-field Quintom model, the scalar fields are
perturbed: $\phi\rightarrow\phi+\delta\phi$ and
$\psi\rightarrow\psi+\delta\psi$. To linear order in the scalar
field fluctuations $\delta \phi$ and $\delta \psi$, the
perturbations of energy, momentum and pressure can be expressed by
\begin{eqnarray}
\label{deltarho}\delta\rho&=&\frac{1}{a^2}\phi'(\delta\phi'-\phi'\Phi)+V_{,\phi}\delta\phi-\frac{1}{a^2}\psi'(\delta\psi'-\psi'\Phi)+W_{,\psi}\delta\psi~,\\
\label{deltaq}\delta q  &=&\frac{1}{a}[-\phi'\delta\phi+\psi'\delta\psi]~,\\
\delta p
&=&\frac{1}{a^2}\phi'(\delta\phi'-\phi'\Phi)-V_{,\phi}\delta\phi-\frac{1}{a^2}\psi'(\delta\psi'-\psi'\Phi)-W_{,\psi}\delta\psi~,
\end{eqnarray}
respectively. The linear terms in the equations obtained by
varying the matter action with respect to the two matter fields,
we obtain the equations of motion for these scalar field
perturbations,
\begin{eqnarray}
\label{peqomf1}{\delta\phi''}&=&-2{\cal H}{\delta\phi'}-k^2\delta\phi-a^2V_{,\phi\phi}\delta\phi-2a^2V_{,\phi}\Phi+4\phi'\Phi'~,\\
\label{peqomf2}{\delta\psi''}&=&-2{\cal
H}{\delta\psi'}-k^2\delta\psi+a^2W_{,\psi\psi}\delta\psi+2a^2W_{,\psi}\Phi+4\psi'\Phi'~.
\end{eqnarray}

\section{A Concrete Example of a Quintom Bounce}

In order to make the analysis more quantitative, in this section
we would like to take a specific example of a Quintom model.
Moreover, our Quintom model should not only provide a bouncing
solution of the universe, but also realize a scenario where the
EoS crosses the cosmological constant boundary and yields a normal
(i.e. non-Phantom-like) evolution of the universe at very early
and very late times. Consequently, we demand that the ghost field
$\psi$ dominates only close to the bounce point. It must be able
to decay after the bounce.

\subsection{Background}

We will assume vanishing potential for the ghost field, whose
background energy density thus consists of kinetic energy
exclusively. From the equation of motion of the ghost field it
follows that $\dot\psi$ evolves proportional to $a^{-3}$. Thus,
the kinetic energy density of the ghost field is proportional to
$a^{-6}$. Therefore, the contribution of the ghost field $\psi$
will dominate at the bounce point, but will decay quickly before
and afterwards. This is just what we need in order to achieve both
a bounce and connect to the observed universe.

We thus  take the Lagrangian of Quintom model to be
\begin{eqnarray}\label{lagrangian}
{\cal
L}=\frac{1}{2}\partial_\mu\phi\partial^\mu\phi-\frac{1}{2}\partial_\mu\psi\partial^\mu\psi-V(\phi)~,
\end{eqnarray}
and, to be  specific, we choose a simple form of the potential
\begin{equation}
V(\phi)=\frac{1}{2}m^2\phi^2 \, .
\end{equation}

\begin{figure}[htbp]
\includegraphics[scale=0.9]{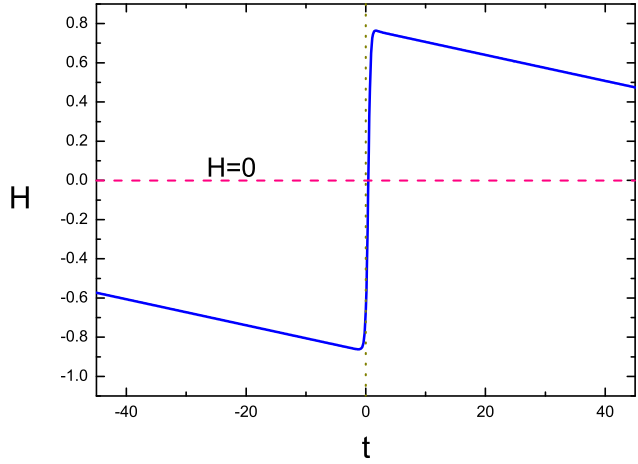}
\caption{Plot of the evolution of the Hubble parameter $H$ in the
model (\ref{lagrangian}). In the numerical calculation we choose
the initial values of parameters as:
$\phi=-5.6\times10^{3},~\dot\phi=2.56\times10^{2},~\dot\psi=4.62\times10^{-73},~m=1.414\times10^{-1}$. The initial time was chosen to be $t = -500$.
Note that in this and the following figures with results from
numerical simulations, all masses are expressed in units of
$10^{-6} M_{pl}$.} \label{fig1:hubb}
\end{figure}

\begin{figure}[htbp]
\includegraphics[scale=0.9]{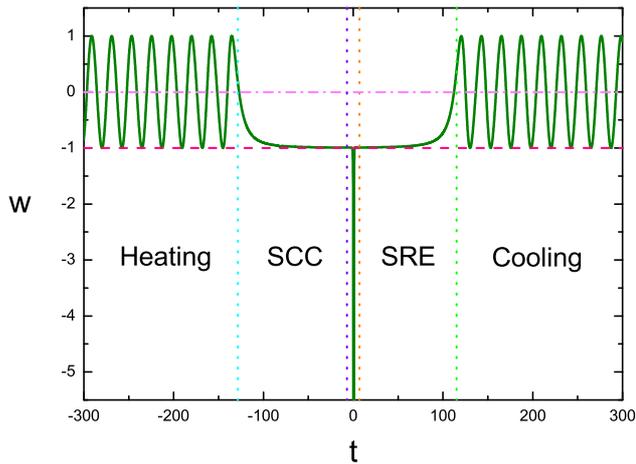}
\caption{Plot of the evolution of the EoS parameter $w$ in the
model (\ref{lagrangian}). The initial values of parameters are the
same as in Fig. \ref{fig1:hubb}.} \label{fig2:eos}
\end{figure}

In this model, the evolution of the universe is symmetric about
the bounce point. In the region long before the bounce point, we
choose initial conditions for which the energy density in the
field $\phi$ dominates by a large factor. Thus, in the initial
phase of evolution (Period 1) the field $\phi$ and hence also the
EoS of the universe will oscillate about $w=0$ and the average
state is similar to that in a matter dominated period. Since the
universe is contracting, it is heating up. Once the amplitude of
the field oscillation gets sufficiently large and the absolute
value of the Hubble expansion parameter increases sufficiently,
the universe will enter a nearly de-Sitter contracting phase
(Period 2) with $w\simeq-1$. This transition is the time reversal
of the end of the slow-rolling period in scalar field inflation.
During this second period, the period of quasi-de-Sitter
contraction, the field $\phi$ climbs slowly up its potential.
During both Periods 1 and 2, but in particular in Period 2, the
contribution of $\psi$ to the energy density is increasing. We
take initial conditions such that at the transition point between
Periods 1 and 2 the contribution of $\psi$ is still subdominant.
Due to the exponential decrease of the scale factor during Period
2, the contribution of $\psi$ rises exponentially and soon starts
to dominate. At this point, the EoS parameter $w$ crosses the
barrier $w =-1$. This triggers the onset of Period 3, the bouncing
phase. The parameter $w$ approaches negative infinity at which
time the bounce occurs. After the bounce, the time reverse of the
evolution before the bounce occurs. The universe again enters a
quasi-de-Sitter phase (Period 4), but this time the universe is
expanding nearly exponentially, i.e., a period of inflation takes
place. During this period, $\phi$ is rolling slowly but the ghost
field decays rapidly, and the EoS returns to a non-Phantom-like
value $w > -1$. Once $\phi$ decreases below a critical value (the
same critical value which signifies the end of inflation in an
$m^2 \phi^2$ inflation model), it will start to oscillate about
the minimum of its potential, and the EoS starts to once again
oscillate around $w=0$.

In Figs. \ref{fig1:hubb} and \ref{fig2:eos} we plot the evolution
of the Hubble parameter and of the EoS, calculated numerically. In
this calculation, we normalized the dimensional parameters such as
$m$, $H$, $\phi$ and $\psi$ by dividing by a mass scale $M$ which
we took to be $10^{-6}M_{pl}$ where the Planck mass is determined
in terms of Newton's gravitational constant by
$M_{pl}=1/\sqrt{G}$. From Fig. \ref{fig2:eos}, we can see that in
this model the universe undergoes a heating phase, a
slow-climb-contracting (SCC) phase, a bounce, a
slow-roll-expanding (SRE) phase, and finally a phase of cooling.
The transitions between the phases are smooth.

Note that the background evolution obtained above is rather generic
and does not depend sensitively on the details of the initial conditions.
The reason is the following:
In our model we have introduced a Phantom field which
makes it rather trivial to obtain a bouncing solution. Since we
have chosen the potential of the Quintom field to vanish, i.e. $W(\psi)=0$,
the Phantom field evolves as
$\dot\psi\propto a^{-3}$ which is similar to matter with its EoS parameter
equal to $1$. Thus, we can conclude that a
bouncing solution is obtained if the absolute value of the
Phantom energy density in the contracting universe grows faster
than that of the normal matter. This will occur if the EoS of the
normal matter is smaller than $1$.In our concrete model
we have chosen normal matter to be a canonical
field $\phi$ with a quadratic potential $V(\phi) = m^2\phi^2/2$,
which leads to an EoS parameter smaller than $1$.
If the initial conditions are imposed during the period
when the Phantom part is negligible, the parameters of the
canonical field such as the mass $m$ and its initial conditions
only affect how long the universe is staying in the heating phase
when $w$ is oscillating around $0$ and in the SCC phase when
($w\simeq-1$).

\subsection{Perturbations}
\label{sec:pertub}

To study the evolution of the perturbation, we combine the Eqs.
(\ref{peqom1}), (\ref{peqom2}) and (\ref{peqom3}) and obtain the
equation of motion of the gravitational potential
\begin{eqnarray}\label{Phieom}
\Phi''+2({\cal H}-\frac{\phi''}{\phi'})\Phi'+2({\cal H}'-{\cal
H}\frac{\phi''}{\phi'})\Phi-\nabla^2\Phi=8\pi G(2{\cal
H}+\frac{\phi''}{\phi'})\psi'\delta\psi~.
\end{eqnarray}

Since there is only a single physical scalar field degree of
freedom for adiabatic scalar metric fluctuations, all other perturbation
variables can be determined from $\Phi$, knowing the evolution of
the background. A frequently used variable is $\zeta$, the curvature
fluctuation in comoving coordinates, which is given by
\begin{eqnarray}
\zeta\equiv\Phi + \frac{{\cal H}}{{\cal H}^2-{\cal H}'}(\Phi'+{\cal
H}\Phi)~.
\end{eqnarray}
In the case of an expanding universe with regular matter, this variable is
known to well describe the adiabatic fluctuations on large scales since
it is conserved on super-Hubble scales because it obeys the equation
$$(1+w)\zeta' \, =  \, \frac{2k^2(\Phi'+{\cal H}\Phi)}{9{\cal H}^2} \, .$$
However, this equation becomes singular both when the universe is
at the bounce point when the Hubble parameter $H$
approaches $0$ and also when the universe is crossing the phantom boundary
$w=-1$. Therefore, $\zeta$ is ill-defined near the bounce
point, as already remarked in \cite{Hassan,Stephon}. Thus, we will
focus on the evolution of $\Phi$ which is well defined throughout
the bounce. We are focusing on adiabatic fluctuations only in this
paper. Since there are two matter fields, it is possible to have
entropy fluctuations. We hope to return to this question in future
work.

\begin{figure}[htbp]
\includegraphics[scale=0.3]{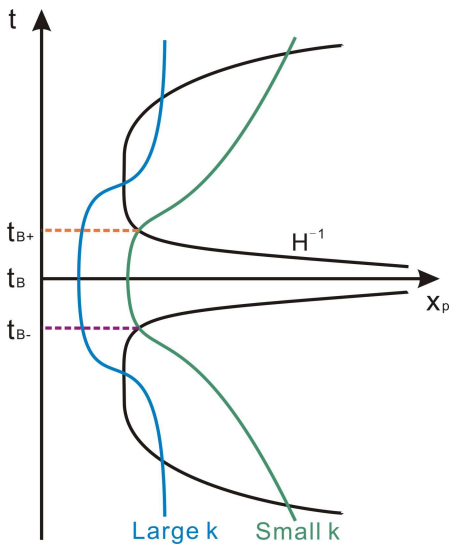}
\caption{A sketch of the evolution of perturbations with different
comoving wave numbers $k$ in the Quintom Bounce.}
\label{fig3:cross}
\end{figure}

Since in the linear theory of cosmological perturbations all
Fourier modes evolve independently, we will follow each mode,
labelled by its comoving wave number $k$, independently.
Interestingly, in our model there are four times when a
perturbation mode crosses the Hubble radius. The evolution of
scales is sketched in Fig. \ref{fig3:cross}. Perturbations start
out inside the Hubble radius in the far past. Since in the initial
heating phase the Hubble radius is shrinking faster than the scale
factor, modes exit the Hubble radius. Once Period 2, the ''slow
climb phase" starts, the scale factor shrinks nearly exponentially
whereas the Hubble radius is constant. Thus, wavelengths which are
not too large will re-enter the Hubble radius during this phase.
During the rest of this phase, and throughout the bounce period,
the scale remains sub-Hubble, and exits the Hubble radius again in
the post-bounce inflationary phase, only to re-enter the Hubble
radius at late times. Modes with an extremely long wavelength will
remain outside the Hubble radius throughout Period 2. Since the
Hubble radius diverges at the bounce point, even these scales will
enter the Hubble radius before the bounce.

To summarize, scales $k$ fall into two broad categories, as
sketched in Fig. \ref{fig3:cross}. The first consists of modes
with comoving wave number $k$ large enough for the perturbations
to return inside the Hubble radius before the bouncing phase and
escape outside after the bouncing phase; the second consists of
modes with small $k$ such that they enter the Hubble radius during
the bouncing phase. We will show later that there is a difference
in the post-bounce spectra of fluctuations between these two range
of scales these.

In a bouncing universe, it is unsuitable to use the equation of
motion for Mukhanov-Sasaki variable $v$ \cite{Mukhanov,Sasaki}
(related closely to $\zeta$)  because this variable is singular
both at the bounce point and the cosmological constant
boundary\cite{Hassan,Stephon,Xia:2007km}. Therefore, in the
following we will directly study the behavior of perturbations
from Eq. (\ref{Phieom}).

In the following we will develop analytical approximations to
study the evolution of the fluctuations in the various phases.
We will denote the time corresponding to the end of inflation
by $t_c$, the end of the bounce phase by $t_B$.

\subsubsection{Heating Phase}

Long before the bounce, when $t \ll -t_B$, we have an average
value of the EoS parameter which is $w=0$, and the contribution of
$\psi$ can be neglected. In this region, we can treat the
evolution of the universe as being dominated by a single field
$\phi$, with the Friedmann equations being
\begin{eqnarray}
{\cal H}^2 \simeq \frac{8\pi G}{3}(\frac{1}{2}\phi'^2+a^2V)~,\\
{\cal H}' \simeq \frac{8\pi G}{3}(-\phi'^2+a^2V)~,
\end{eqnarray}
using conformal time and Hubble parameter. Further we have
\begin{equation}
\frac{\phi''}{\phi'}=\frac{2{\cal H}{\cal H}'-{\cal H}''}{2({\cal
H}^2-{\cal H}')} \, ,
\end{equation}
and $a\propto\eta^2$, with $\eta < 0$ during the contracting phase.
Thus, we obtain the expressions
\begin{equation}
{\cal H} = \frac{2}{\eta}, \,\,\,\
{\cal H}' = -\frac{2}{\eta^2} \,\,\,\
{\cal H}'' = \frac{4}{\eta^3} \, .
\end{equation}

Making use of these relations, we obtain the following equation
for the gravitational potential
\begin{eqnarray}
\Phi_k''+\frac{6}{\eta}\Phi_k'+k^2\Phi_k=0~,
\end{eqnarray}
in momentum space. Note that this equation is very similar to the
perturbation equation in a matter dominant period except that the
last term $k^2\Phi_k$ does not exist for hydrodynamical
perturbations (ordinary hydrodynamical matter is non-relativistic,
our scalar field matter is relativistic). The analytical solution
of the above equation of motion can be written as follows,
\begin{eqnarray}
\Phi_k=\eta^{-\frac{5}{2}}[k^{-\frac{5}{2}}A_k
J_{\frac{5}{2}}(k\eta)+k^{\frac{5}{2}}B_k
J_{-\frac{5}{2}}(k\eta)]~,
\end{eqnarray}
where the coefficients $A_k$ and $B_k$ can be determined from the
initial condition.

To set the initial conditions, we first define the variable
\begin{equation}
u_k\equiv\frac{a\Phi_k}{4\pi G\phi'}\, .
\end{equation}
This is the variable in terms of which the initial conditions are
determined by canonical quantization \cite{MFB}. We will take the
Bunch-Davis vacuum as the initial conditions (when $\eta$ is
sufficiently far away from the bounce point) \footnote{We will
discuss other possible initial conditions in a followup paper.}:
\begin{equation}
u_k=\frac{1}{\sqrt{2k^3}}e^{-ik\eta} \, .
\end{equation}
During Period 1 when $w=0$ we  have
$\frac{\phi'}{a}\sim\eta^{-3}$. Therefore the initial condition
for the gravitational potential becomes
\begin{eqnarray}
\Phi_k = A \eta^{-3} \frac{e^{-ik\eta}}{\sqrt{2k^3}}~~,~~A\equiv
4\pi G\sqrt{\rho_0}\eta_0^3~,
\end{eqnarray}
where $A$ is a normalization factor. When $|k\eta|\gg1$ we can
take the asymptotical form of the Bessel function and, matching
with the Bunch-Davies vacuum, determine the coefficients. The
result is
\begin{eqnarray}
A_k=i\frac{\sqrt{\pi}}{2}Ak^{\frac{3}{2}}~~,~~
B_k=-\frac{\sqrt{\pi}}{2}Ak^{-\frac{7}{2}}~.
\end{eqnarray}
When the wavelength of perturbation is larger than Hubble radius
with $|k|\ll {\cal H}$, then by using another asymptotic form of
the Bessel functions we obtain the following form of $\Phi_k$:
\begin{eqnarray}\label{heating}
\Phi_k\simeq\frac{iAk^{3/2}}{15\sqrt{2}}-\frac{3A}{\sqrt{2}}k^{-7/2}\eta^{-5}~.
\end{eqnarray}
From this result, we can see that of the two modes of $\Phi$ one
is constant and the other growing during the contracting heating
phase.

\subsubsection{Slow-Climb-Contracting (SCC) Phase}
\label{sec:SCC}

For the universe in the contracting phase, we have $H<0$. In this
case, $3H\dot\phi$ is anti-frictional, and instead of damping the
motion of $\phi$ in the expanding phase it accelerates the motion
of $\phi$. Consequently, the canonical scalar field $\phi$ will
eventually stop oscillating and start to climb up its potential.
We denote this transition time as $\eta_c$. For ($\eta > \eta_c$)
on, the universe is in a period of SCC phase with $w\simeq-1$. We
assume that in the beginning of this period the ghost field can
still be neglected. Therefore, the equation of motion for the
gravitational potential becomes
\begin{eqnarray}\label{Phieomsc}
\Phi_k''+2\sigma{\cal H}\Phi_k'+2(\sigma-\epsilon){\cal
H}^2\Phi_k+k^2\Phi_k=0~,
\end{eqnarray}
where we take the slow-climb(slow-roll) parameters as
\begin{equation}
\epsilon\equiv-\frac{\dot H}{H^2} \,\,\, {\rm and} \,\,\,
\sigma\equiv-\frac{\ddot H}{2H\dot H} \,~.
\end{equation}

As is well known, we have
\begin{equation}
\eta - \eta_c \simeq 1/{\cal H}_c - 1/{\cal H} \,~,
\end{equation}
where a subscript `c' indicates that the corresponding quantity is
evaluated at the time $\eta_c$.

Making use of the above relations, we can solve the equation
(\ref{Phieomsc}) for the gravitational perturbation:
\begin{eqnarray}\label{Phisc}
\Phi_k=(\eta-\tilde{\eta}_c)^{\alpha}[k^{-\nu}C_k
J_{\nu}(k(\eta-\tilde{\eta}_c))+k^{\nu}D_k
J_{-\nu}(k(\eta-\tilde{\eta}_c))]~,
\end{eqnarray}
where $\alpha\simeq\frac{1}{2}$, $\nu\simeq\frac{1}{2}$ and
$\tilde{\eta}_c\equiv\eta_c+1/{\cal H}_c$. We can obtain the
asymptotic form of Eq. (\ref{Phisc}) in the sub-Hubble region
\begin{equation}
\Phi_k \simeq
\sqrt{\frac{2}{\pi}}[\frac{C_k}{k}\sin(k(\eta-\tilde\eta_c))+D_k\cos(k(\eta-\tilde\eta_c))]~,
\end{equation}
and in the super-Hubble region
\begin{equation}
\Phi_k \simeq \sqrt{\frac{2}{\pi}}[C_k(\eta-\tilde{\eta}_c)+{D_k}]
\,~.
\end{equation}
In the latter form the two modes are a constant mode $D_k$ and a
mode with coefficient $C_k$ which is growing.

Through matching $\Phi$ and $\zeta$ (these are the matching
conditions derived in \cite{Hwang,Deruelle}) in the heating phase
and in the SCC phase at the time $\eta_c$, we can determine the
coefficients $C_k$ and $D_k$ as follows,
\begin{eqnarray}
C_k&=&-{\cal H}_c\left[\frac{1}{15}(1-\frac{2}{3}\epsilon)A_k+3(1+\epsilon)B_k\eta_c^{-5}\right]~,\nonumber\\
D_k&=&\epsilon(\frac{2}{45}A_k-3B_k\eta_c^{-5})\simeq0~.
\end{eqnarray}
Therefore, in the SCC phase the main contribution to the
perturbation comes from the initial growing mode of $\Phi_k$.

\subsubsection{Bouncing Phase}

Once the universe enters the quasi-de-Sitter contracting phase,
the contribution of the ghost field $\psi$ becomes more and more
important, since its kinetic energy density scales as $a^{-6}$.
Therefore, the SCC phase will cease at some moment $\eta_{B-}$
when the contribution of $\psi$ begins to dominate, and then the
EoS parameter of the universe will cross $w = -1$ and fall to
negative infinity rapidly. Correspondingly, the energy density and
the Hubble parameter reach zero when the bounce takes place at the
time $\eta_B$. Since at this moment the Hubble radius $1/H$
approaches infinity, all wavelengths will be inside the Hubble
radius. Thus, the perturbations will be once again oscillating
close to the bounce point. Short wavelength modes will have time
to perform many oscillations which are sub-Hubble near the bounce
point, whereas long wavelength modes (which have a longer
intrinsic period and also spend less time being sub-Hubble) will
not have time to perform a complete oscillation. As we will see,
this difference leads to a difference in the way the initial
spectrum of fluctuations transfers through the bounce.

It is rather complicated to solve the perturbation equation
exactly from Eq. (\ref{Phieom}). In order to gain some analytical
insight into the evolution of fluctuations in this region, we
resort to some approximations in order to simplify the equation.

A convenient parametrization of the Hubble parameter near the
bounce point (we choose $t = 0$ to correspond to the bounce point)
is
\begin{equation} \label{param}
H(t)  = \alpha t \, ,
\end{equation}
where $\alpha$ is a positive constant.

In this case, we obtain the analytic forms of the scale factor and
the comoving Hubble parameter:
\begin{eqnarray}
a=\frac{a_B}{1-\frac{2}{\pi}\alpha a_B^2 (\eta-\eta_B)^2}~,~{\cal
H}=\frac{\frac{4}{\pi}\alpha
a_B^2(\eta-\eta_B)}{1-\frac{2}{\pi}\alpha a_B^2(\eta-\eta_B)^2}~,
\end{eqnarray}
where $a_B$ denotes the scale factor at the bounce point.

In the bouncing process, the field $\phi$ is initially climbing
up to the hill. However, as the absolute value of $H$ is
decreasing, $\ddot\phi$ changes its sign which then leads to
the field $\phi$ reversing its direction and beginning  to slowly
roll down the potential. At this stage, the universe has bounced
and $H$ is positive.

\begin{figure}[htbp]
\includegraphics[scale=1.0]{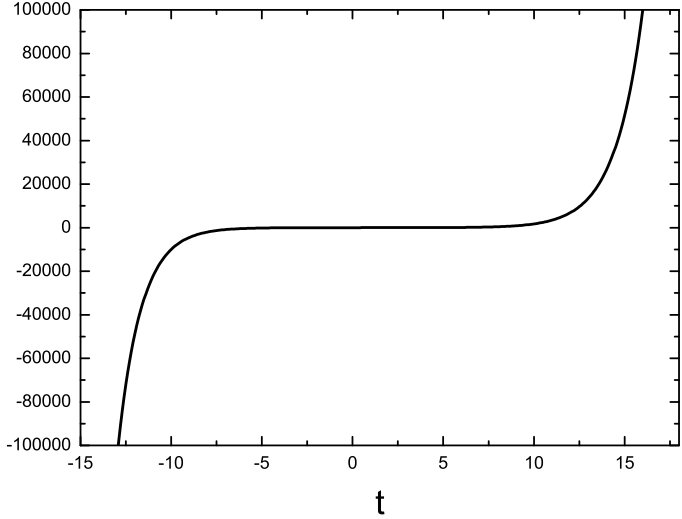}
\caption{A plot of the evolution of the factor
$3H+\frac{\ddot\phi}{\dot\phi}$ in the bouncing phase.}
\label{fig35:prefactor}
\end{figure}

Making use of the parameterization $H = \alpha t$
(see (\ref{param})) we can solve Eq. (\ref{doyHbg}) to obtain
\begin{eqnarray}
\frac{\ddot\phi}{\dot\phi}=-3\alpha t
\dot\psi^2/(\dot\psi^2-\frac{\alpha}{4\pi G})\simeq-3H~,
\end{eqnarray}
around the bounce point. Consequently, we can derive another
formula $\phi''\simeq-2{\cal H}\phi'$. In Sec. \ref{sec:nr}, we
will show that this is a very successful simulation by comparing
with the numerical calculation. In order to show this point more
explicitly, in Fig. \ref{fig35:prefactor} we present a plot of the
factor $3H+\frac{\ddot\phi}{\dot\phi}$ in the bouncing phase. The
values shown in the figure
support the approximation we have made in Eq. (42) since from the figure,
it can be seen that this factor vanishes around the bounce point.

We suppose that the approximations above should be valid only in
the neighborhood of the bounce point, and hence the square term
and higher order terms of $|\eta-\eta_B|$ should be neglected.

Substituting these approximations into Eq. (\ref{Phieom}), we
obtain the following equation:
\begin{eqnarray}
\Phi_k''+3y_1(\eta-\eta_B)\Phi_k'+(k^2+y_1)\Phi_k\simeq 0~,
\end{eqnarray}
where we define the parameter $y_1\equiv\frac{8}{\pi}\alpha
a_B^2$. The solution to this equation can be written as
\begin{eqnarray}\label{Phibbs}
\Phi_k = e^{-\frac{3}{2}y_1(\eta-\eta_B)^2} \left[E_k {\rm
H}_{\bar{m}}(\sqrt{3y_1/2}(\eta-\eta_B)) +
F_k~_1F_1(-\frac{\bar{m}}{2};\frac{1}{2};\frac{3}{2}y_1(\eta-\eta_B)^2)
\right]~,
\end{eqnarray}
which is constructed by a $\bar{m}$-th Hermite polynomial and a
confluent hyper-geometric function with
$\bar{m}=\frac{k^2-2y_1}{3y_1}$ and two undetermined coefficients
$E_k$, $F_k$. These two functions are linearly independent, but
both of them show oscillating behavior when $y_1/k^2$ is
sufficiently small. This is to be expected in our bounce phase
since the Hubble radius blows up at the bounce point and all
scales become sub-Hubble and hence the associated mode functions
should oscillate, even for $|k(\eta-\eta_B)|\ll1$.

\begin{figure}[htbp]
\includegraphics[scale=1.0]{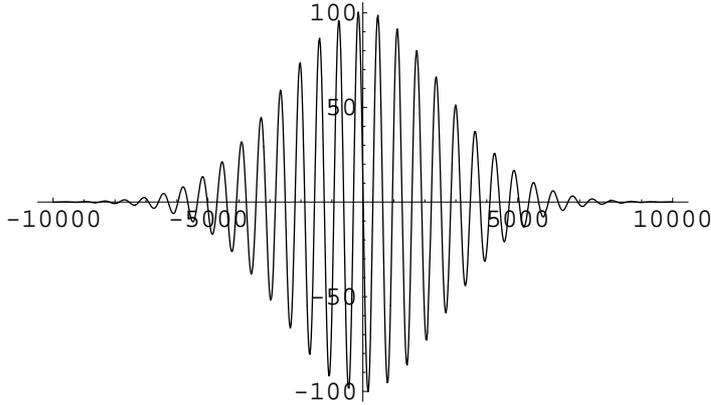}
\caption{A sketch of the evolution of the gravitational potential
$\Phi$ in the bouncing phase.} \label{fig4:osc}
\end{figure}

If the parameter $\bar{m}$ is big enough, We can make use of the
following approximate
\begin{eqnarray}\label{bouncing}
\Phi_k\simeq e^{-\frac{3}{4}y_1(\eta-\eta_B)^2}\left\{ \tilde
F_k\cos[k(\eta-\eta_B)] + \tilde E_k\sin[k(\eta-\eta_B)]
\right\}~,
\end{eqnarray}
where
\begin{equation}
\tilde E_k \equiv
-\frac{2^{\bar{m}+1}\sqrt{\pi}}{\sqrt{2\bar{m}}\Gamma(-\frac{\bar{m}}{2})}E_k
\end{equation}
and
\begin{equation}
\tilde F_k \equiv
\frac{2^{\bar m}\sqrt{\pi}}{\Gamma(\frac{1-\bar{m}}{2})}E_k+F_k \, .
\end{equation}
Interestingly, we obtain a factor
$e^{-\frac{3}{4}y_1(\eta-\eta_B)^2}$ which can enlarge the
amplitude of the perturbation before the bounce and suppress it
after the bounce. We sketch this solution in Fig. \ref{fig4:osc}
and will find that it is consistent with the numerical results in
Sec. \ref{sec:nr}.

Note that there are two possible ways for the modes in the SCC
phase to evolve into the bouncing phase, as described in the
beginning of Sec. \ref{sec:pertub} (also see Fig.
\ref{fig3:cross}). The first possibility is that the comoving wave
number $k$ is large enough such that the mode reenters the Hubble
radius in the SCC phase. The second is that $k$ is small enough
and the mode re-enters the Hubble radius only during the bouncing
phase.

In the first case with $k\gg{\cal H}_{B-}$, the spectrum has been
oscillating already at late times in the SCC phase. Through
matching $\Phi$ and $\zeta$ in the SCC phase and in the bouncing
phase at the time $\eta_{B-}$ we can determine the coefficients
$\tilde E_k^{(1)}$ and $\tilde F_k^{(1)}$ as follows,
\begin{eqnarray}
 \tilde E_k^{(1)} &\simeq&
e^{\frac{3}{4}y_1(\eta_{B-}-\eta_B)^2}\sqrt{\frac{2}{\pi}} \left\{
(\frac{C_k}{k}\sin[k(\eta_{B-}-\tilde\eta_c)]+D_k\cos[k(\eta_{B-}-\tilde\eta_c)])\sin[k(\eta_{B-}-\eta_B)]\right.
\nonumber\\
&&\left. -\frac{1}{k(\eta_{B-}-\eta_B)}\frac{1}{\epsilon{\cal
H}_{B-}}[(C_k+{\cal
H}_{B-}D_k)\cos[k(\eta_{B-}-\tilde\eta_c)]\right.\nonumber\\
&&\left. +(\frac{{\cal
H}_{B-}}{k}C_k-kD_k)\sin[k(\eta_{B-}-\tilde\eta_c)]]\cos[k(\eta_{B-}-\eta_B)]
\right\}~,\\
 \tilde F_k^{(1)} &\simeq&
e^{\frac{3}{4}y_1(\eta_{B-}-\eta_B)^2}\sqrt{\frac{2}{\pi}} \left\{
(\frac{C_k}{k}\sin[k(\eta_{B-}-\tilde\eta_c)]+D_k\cos[k(\eta_{B-}-\tilde\eta_c)])\cos[k(\eta_{B-}-\eta_B)]\right.
\nonumber\\
&&\left. +\frac{1}{k(\eta_{B-}-\eta_B)}\frac{1}{\epsilon{\cal
H}_{B-}}[(C_k+{\cal
H}_{B-}D_k)\cos[k(\eta_{B-}-\tilde\eta_c)]\right.\nonumber\\
&&\left. +(\frac{{\cal
H}_{B-}}{k}C_k-kD_k)\sin[k(\eta_{B-}-\tilde\eta_c)]]\sin[k(\eta_{B-}-\eta_B)]
\right\}~.
\end{eqnarray}
From this result we can read off that, in the case when $k\gg{\cal
H}_{B-}$ the coefficients of the perturbations in bouncing phase
depend in an UN-suppressed way on the coefficient $C_k$ of the
growing mode in the SCC phase. However, this is not an absolute
relation since the coefficient $D_k$ of the constant mode in the
SCC phase may play an important role when we fine-tune the phases
$k(\eta_{B-}-\eta_c)$ and $k(\eta_{B-}-\eta_B)$.

Secondly, we consider the case when $k\sim{\cal H}_{B-}$. Through
matching the spectrum in the bouncing phase with that in the SCC
phase, the coefficients can be determined as follows,
\begin{eqnarray}
 \tilde E_k^{(2)} &\simeq&
e^{\frac{3}{4}y_1(\eta_{B-}-\eta_B)^2}\sqrt{\frac{2}{\pi}} \left\{
(D_k-\frac{C_k}{{\cal H}_{B-}})\sin[k(\eta_{B-}-\eta_B)]\right.
\nonumber\\
&&\left.
-\frac{1}{k(\eta_{B-}-\eta_B)}\frac{D_k}{\epsilon}\cos[k(\eta_{B-}-\eta_B)]\right\}\nonumber\\
&\simeq&-e^{\frac{3}{4}y_1(\eta_{B-}-\eta_B)^2}\frac{\sqrt{2/\pi}}{k(\eta_{B-}-\eta_B)}\frac{D_k}{\epsilon}\cos[k(\eta_{B-}-\eta_B)]~,\\
 \tilde F_k^{(2)} &\simeq&
e^{\frac{3}{4}y_1(\eta_{B-}-\eta_B)^2}\sqrt{\frac{2}{\pi}} \left\{
(D_k-\frac{C_k}{{\cal H}_{B-}})\cos[k(\eta_{B-}-\eta_B)]\right.
\nonumber\\
&&\left.
+\frac{1}{k(\eta_{B-}-\eta_B)}\frac{D_k}{\epsilon}\sin[k(\eta_{B-}-\eta_B)]\right\}\nonumber\\
&\simeq&e^{\frac{3}{4}y_1(\eta_{B-}-\eta_B)^2}\frac{\sqrt{2/\pi}}{k(\eta_{B-}-\eta_B)}\frac{D_k}{\epsilon}\sin[k(\eta_{B-}-\eta_B)]~.
\end{eqnarray}
From these two equations we find that it is usually the
coefficient $D_k$ of the sub-dominant mode in the SCC phase which
mainly dominates the coefficients of the perturbations in the
bouncing phase, unless we fine-tune the phase
$k(\eta_{B-}-\eta_B)$.

\subsubsection{Slow-Roll-Expanding Phase (SRE)}

Due to symmetry of the Quintom model, the scalar field $\phi$
begins to slow roll down to the bottom of its potential and the
ghost field $\psi$ decays soon after the bounce. Therefore, a
phase of slow-roll expansion (like slow-roll inflation) starts at
the time $\eta_{B+}$. We denote the scale factor at this moment as
$a_{B+}$. During this period, the Hubble parameter is nearly
constant which is similar to what occurs in the usual inflationary
scenario. Interestingly, the EoS of the universe evolves from
below $-1$ to above which provides a link with the normal
expanding history of our universe.

The equation of motion for the gravitational potential in this phase is
the same as in the SCC phase (Eq. (\ref{Phieomsc})). Thus, the
solutions takes the form
\begin{eqnarray}\label{Phisre}
\Phi_k=(\eta-\tilde{\eta}_{B+})^{\alpha}[k^{-\nu}G_kJ_\nu(k(\eta-\tilde{\eta}_{B+}))+k^{\nu}H_kJ_{-\nu}(k(\eta-\tilde{\eta}_{B+}))]~,
\end{eqnarray}
where $\tilde\eta_{B+}\equiv\eta_{B+}+1/{\cal H}_{B+}$ and the
values of $\alpha$ and $\nu$ are the same as those which appeared
in Sec. \ref{sec:SCC}. We can obtain the asymptotical form of Eq.
(\ref{Phisre}) in the sub-Hubble region
\begin{equation}
\Phi_k\simeq\sqrt{\frac{2}{\pi}}[\frac{G_k}{k}\sin(k(\eta-\tilde\eta_{B+}))+H_k\cos(k(\eta-\tilde\eta_{B+}))]
\end{equation}
and in the super-Hubble region
\begin{equation}
\Phi_k\simeq \sqrt{\frac{2}{\pi}}[G_k(\eta-\tilde{\eta}_{B+})+{H_k}] \, .
\end{equation}
The second mode is constant (it corresponds to the usual dominant
mode on super-Hubble scales in an expanding universe)
and the first mode is decreasing.

When determining the coefficients of the two perturbation modes,
we again need to match with the perturbation in bouncing phase in
two cases. In the case $k\gg{\cal H}_{B+}$ and assuming that all
the phases are nontrivial, we obtain
\begin{eqnarray}
G_k^{(1)} &\simeq&
e^{-\frac{3}{4}y_1(\eta_{B+}-\eta_B)^2}\frac{\sqrt{\pi/2}}{2\sin[k(\eta_{B+}-\tilde
\eta_{B+})]} k \left\{\tilde E_k\sin[k(\eta_{B+}-\eta_B)]+\tilde
F_k\cos[k(\eta_{B+}-\eta_B)]\right\}~,\\
H_k^{(1)} &\simeq&
e^{-\frac{3}{4}y_1(\eta_{B+}-\eta_B)^2}\frac{\sqrt{\pi/2}}{2\cos[k(\eta_{B+}-\tilde
\eta_{B+})]} \left\{\tilde E_k\sin[k(\eta_{B+}-\eta_B)]+\tilde
F_k\cos[k(\eta_{B+}-\eta_B)]\right\}~.
\end{eqnarray}
Therefore, in this case both coefficients are of the same order
which means that the two modes are both important in the
sub-Hubble region. However, as soon as the perturbation evolves
outside the Hubble radius, the decreasing mode becomes negligible.
The amplitude of the dominant mode in the expanding phase is set
by the amplitude of the growing mode in the contracting phase. We
will compare our result with that obtained in other works on
non-singular bouncing cosmologies in the Discussion section.

In the other case we take $k\sim{\cal H}_{B+}$ and also assume all
the phases are nontrivial. We then obtain
\begin{eqnarray}
G_k^{(2)} &\simeq&-e^{-\frac{3}{4}y_1(\eta_{B+}-\eta_B)^2}{\cal
H}_{B+}\sqrt{\frac{\pi}{2}} \left\{\tilde
E_k\sin[k(\eta_{B+}-\eta_B)]+\tilde
F_k\cos[k(\eta_{B+}-\eta_{B})]\right\}~,\\
H_k^{(2)}
&\simeq&-e^{-\frac{3}{4}y_1(\eta_{B+}-\eta_B)^2}(\eta_{B+}-\eta_B)\epsilon
\sqrt{\frac{\pi}{2}}k\left\{\tilde
E_k\cos[k(\eta_{B+}-\eta_B)]-\tilde
F_k\sin[k(\eta_{B+}-\eta_{B})]\right\} \nonumber\\
&\simeq& 0~.
\end{eqnarray}
Thus, we obtain the result that for these very long wavelength
mode, the perturbations contain, to leading order in $k$, only the
decreasing mode at late time. This is similar to the behavior of
fluctuations in models with a singular bounce, such as the
Pre-Big-Bang \cite{PBB} and Ekpyrotic universe \cite{Ekp}
scenarios. In the Pre-Big-Bang scenario, it has been shown
\cite{BGGMV} that the growing mode of the fluctuations in the
contracting phase matches (up to correction factors which are
suppressed by powers of $k$) only to the decaying mode in the
expanding phase. The amplitude of the dominant mode in the
expanding phase is thus set by the amplitude of the constant (and
thus sub-dominant) mode in the contracting phase. The same result
also holds \cite{Lyth,Fabio,Hwang2,Creminelli} in the case of four
dimensional toy model descriptions of the Ekpyrotic scenario,
although in the case of the Ekpyrotic scenario the physical model
is defined in higher dimensions, and an analysis of the evolution
of fluctuations in the higher dimensional context
\cite{Tolley,Thorsten} shows that there is a non-trivial mixing
between the growing mode in the contracting phase and the dominant
mode in the expanding phase.

\subsubsection{Numerical Results}\label{sec:nr}

\begin{figure}[htbp]
\includegraphics[scale=0.9]{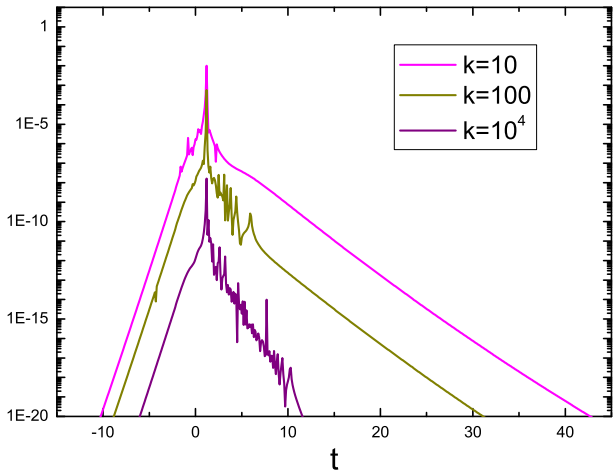}
\caption{Plot of the evolution of the absolute value of the ratio
of the r.h.s to the term $k^2\Phi$ in Eq. (\ref{Phieom}). The
initial values of the background parameters are the same as in
Fig. \ref{fig1:hubb}.} \label{fig45:ratio}
\end{figure}

In the above sections we have present the analytic calculation of
our model, during the process we have taken an assumption that the
right hand side of Eq. (\ref{Phieom}) is negligible during the
whole evolution. This is a key to solving the perturbation
equation in our model explicitly. As is pointed out in previous
sections, the energy density of the phantom field is usually
enough small in the heating phase, the SCC phase, the SRE phase
and the cooling phase and hence we have $\psi'\simeq 0$. Therefore
it is a good approximation to neglect the right hand side of Eq.
(\ref{Phieom}) in these four phases. Moreover, we have argued that
the relation $2{\cal H}+\frac{\phi''}{\phi'} \simeq 0$ in the
bouncing phase which makes the right hand side of Eq.
(\ref{Phieom}) negligible. To make sure the assumption valid
during the whole evolution, we plot the ratio of the r.h.s to the
term $k^2\Phi$ appearing in the equation as shown in Fig.
\ref{fig45:ratio}. From this figure we can find that the value of
the ratio in our model is always much less than $1$ which agrees
with our assumption very well.

\begin{figure}[htbp]
\includegraphics[scale=0.9]{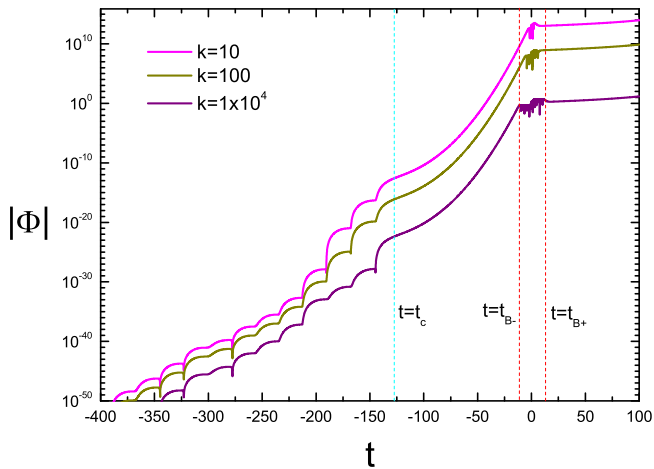}
\caption{Plot of the evolution of the absolute value of the
gravitational potential $|\Phi|$ in our model. The initial values
of the background parameters are the same as in Fig.
\ref{fig1:hubb}.}
\label{fig4:absPhi}
\end{figure}

\begin{figure}[htbp]
\includegraphics[scale=0.9]{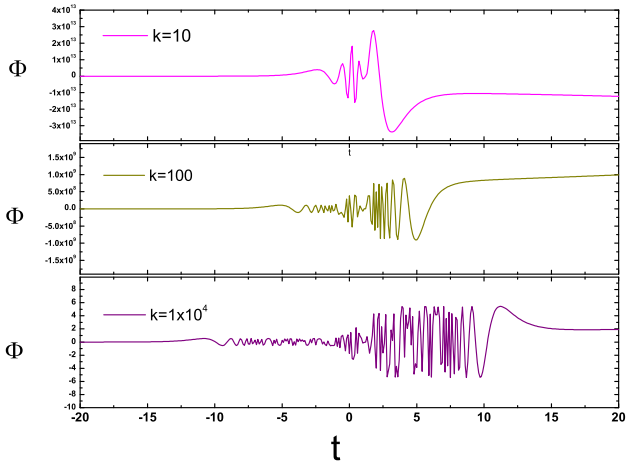}
\caption{Plot of the evolution of the gravitational potential
$\Phi$ in our model. The initial values of the background
parameters are the same as in Fig. \ref{fig1:hubb}.}
\label{fig5:Phi}
\end{figure}

\begin{figure}[htbp]
\includegraphics[scale=0.9]{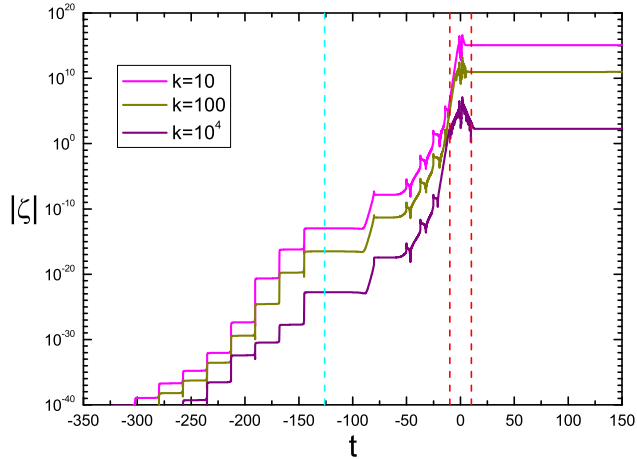}
\caption{Plot of the evolution of the absolute value of the
curvature perturbation $|\zeta|$ in our model. The initial values
of the background parameters are the same as in Fig.
\ref{fig1:hubb}.} \label{fig55:zeta}
\end{figure}

In order to make the previous analysis exact and precise, we also
solved the perturbation equations numerically. We provide the
numerical result in Fig. \ref{fig4:absPhi}, Fig. \ref{fig5:Phi},
and Fig. \ref{fig55:zeta}. In Fig. \ref{fig4:absPhi}, we plot the
absolute value of the Newtonian gravitational potential $|\Phi|$
for different values of $k$. We can see from this figure that in
the period far before the bounce point, those curves are similar.
At the beginning of the evolution, when the wavelength is
sub-Hubble, the perturbation is near its Bunch-Davies vacuum and
its value oscillates with increasing amplitude. One may notice
that the initial oscillatory behaviors of different modes are the
same as others which only depend on the background evolution in
Fig. \ref{fig4:absPhi}. We give the explanations as follows. In
our model, the canonical field $\phi$ initially oscillates around
its vacuum in the contracting period which results in a heating
phase. During this phase, the EoS oscillates between $-1$ and $1$
as shown in Fig. \ref{fig2:eos}. In our analytical calculation we
simply take the average value of the EoS $w=0$ and show that there
are two modes for the gravitational potential with one being
constant and the other being growing mode. However, when we take
the numerical calculation, we encounter the case that there has to
be a point satisfying $\dot\phi=0$ and $w=-1$ in every
oscillation. At this moment, there can indeed be a large growth of
fluctuations due to parametric resonance. Therefore it seems that
there are oscillations on the evolution of the gravitational
potential, but it is not true. Since these resonances are
determined by the background evolution, those ``oscillations" have
the same frequency and features regardless of the wave number $k$.
By the same reason we can understand the periodical amplifications
of $|\zeta|$ as they appear in Fig. \ref{fig55:zeta}. This issue
will be discussed in future work. After the mode exits the Hubble
radius, the perturbation will be dominated by its growing mode, as
is shown in Eq. (\ref{heating}). When it enters into the SCC
phase, it grows monotonically at first and becomes oscillatory
near the bounce point after the wavelength re-enters the Hubble
radius.

During the bouncing phase, the curves of different $k$-modes
behave very differently. For large $k$, the oscillatory behavior
starts earlier and lasts longer because the wavelength of such
modes enters into the sub-Hubble region much earlier before the
bounce and escapes from the Hubble radius much later after the
bounce. The amplitude of the oscillations near the bounce point
increases before the bounce and decreases after the bounce, which
follows from the Hubble term in the equation of motion which acts
as anti-friction in the contracting phase and friction in the
expanding phase, as shown in Eq. (\ref{bouncing}). On the other
hand, the oscillating period for small $k$-modes becomes much
shorter since these modes do not enter in the sub-Hubble region
until after the bouncing phase has started.

Finally, after the bounce the universe enters into the SRE phase
and the wavelengths of the fluctuations become super-Hubble once
again. Now, there are also two very different evolution scenarios
for fluctuations with different $k$-modes. In the expanding phase,
for large enough values of $k$, the gravitational potential is
dominated by the constant mode; while in the case of small values
of $k$, it is dominated by decreasing mode.

We also plot the values of $\Phi$ of different modes around the
bounce point in Fig. \ref{fig5:Phi}. We can see that, for larger
values of $k$, the oscillations begins much earlier with their
amplitude being enlarged at the bounce point, whereas for smaller
values of $k$ there is only a very short period of oscillation
(with amplified amplitude) which is caused by bounce.

After the bounce, the gravitational potential for large $k$
evolves almost as a constant. However, for small $k$ there is a
decrease on the gravitational potential. All these features are
consistent with the analytical results discussed above.

Finally, we also provide a plot of the curvature perturbation
$\zeta$ in Fig. \ref{fig55:zeta}.
From this figure we can see that there inevitably are oscillations of the
perturbations near the bounce point.

\section{Discussion And Conclusions}

Bouncing non-singular cosmologies, with an initial contracting
phase which lasts until to a non-vanishing minimal radius is
reached and then smoothly transits into an expanding phase,
provide a possible solution to the singularity problem of Standard
Big Bang cosmology, a problem which is not cured by
scalar-field-driven inflationary models.

In this paper, we have considered the evolution of cosmological
fluctuations in a particular bouncing cosmology, namely the
Quintom Bounce. In this model, space-time continues to be
described by Einstein's General Relativity. The violations of the
null energy condition required to get a bounce are obtained by
making use of Quintom matter, which allows a transition of the EoS
$w$ through the cosmological constant boundary. We have studied
the evolution of the gravitational potential both analytically and
numerically in a particular model of a Quintom Bounce in which
matter consists of two scalar fields, one of them a
Quintessence-like field, the other a Phantom-like field. We have
found that the evolution of the gravitational potential $\Phi$ is
completely non-singular. Whereas in models with a singular bounce
such as the Pre-Big-Bang \cite{PBB} and the Ekpyrotic \cite{Ekp}
scenarios, the perturbations must be matched across a singular
space-like surface, which leads to serious conceptual problems
\cite{Durrer}, in our case no such problems arise
\footnote{Recently, a ghost condensation mechanism to smooth out
the singularity in the Ekpyrotic scenario has been introduced
\cite{Senatore,Khoury}, and it has been shown in this context,
using the techniques developed in \cite{Riotto,Fabio3,Turok} that
making use of iso-curvature modes, the dominant spectrum in the
contracting phase can be passed on to the dominant spectrum in the
expanding phase.}. In an analytical approximation scheme, we can
apply matching conditions at non-singular surfaces which are then
well justified, and we can also study the evolution numerically
without any approximations.

Our second main result is that for short wavelength modes, modes
which re-enter the Hubble radius in the slow contracting phase,
the dominant fluctuation mode is determined by the growing mode in
the contracting phase. This result is similar to what has recently
been found in other non-singular bouncing cosmologies
\cite{Hassan,Stephon,Abramo}. The first of these is a bounce in
the context of mirage cosmology, the second in the context of a
specific higher derivative gravity model, and the third in the
context of K-essence. Earlier work on fluctuations in models with
a kinetic term with opposite sign generating a nonsingular bounce
is in \cite{Peter}. The authors of \cite{Peter} observed that the
way the post-bounce spectrum depends on the pre-bounce spectrum
depends sensitively on the details of the bounce. In this light,
it is not surprising that other bounces show that the dominant
mode in the pre-bounce phase does not couple to the dominant mode
in the post-bounce phase
\cite{Cartier,Tsujikawa,Bozza,Wands,Finelli2,Veneziano}. For long
wavelength modes, however, we find that the dominant fluctuation
mode in the contracting phase at leading order in $k$ only sources
the decaying mode in the expanding phase, a result similar to what
is obtained in singular bouncing models
\cite{BGGMV,Lyth,Fabio,Hwang2,Creminelli}.

Thus, the fluctuations in the Quintom Bounce model evolve in way
which combines aspects found in some other recently studied
non-singular bouncing cosmologies \cite{Hassan,Stephon} (the
behavior of short wavelength modes) with aspects found in singular
bouncing cosmologies (the behavior of the long wavelength modes).
Which category of modes determines scales accessible to current
cosmological observations depends on the detailed parameters of
the model. Our model may help us to understand the physics of
singular bounce and non-singular bounce more clearly. As a side
remark, let us add that in the Quintom Bounce model there is no
trans-Planckian problem \cite{RHBrev2,Martin} for fluctuations
since, unless the periods of slow-climb contraction and slow-roll
expansion last many Hubble expansion times, the wavelengths of all
the observable perturbations will be larger than the Planck length
at the bounce point and thus never enter the zone of
trans-Planckian ignorance.

Let us close with some more general comments related to the
Quintom Bounce. Although this theory is still in the stage of
development, there have been works done in the context of a
bouncing scenario related to predictions for observations. For
example, in \cite{Piao:2003zm} (see also
\cite{Piao:2003hh,Piao:2005ag}) it was pointed out that with a
bounce followed by slow-roll inflation it is possible to give a
reasonable explanation for the suppression of the low multi-poles
of the CMB anisotropies. In \cite{Cai:2007qw} and
\cite{Aref'eva:2007uk} the possibility of obtaining a bounce in a
Quintom model with a single matter field with higher-derivative
terms was explored. See also \cite{Zhang:2007bi,Wei:2007rp} for
other works.

\section*{Acknowledgments}

It is a pleasure to thank Mingzhe Li, Jun-Qing Xia and Yi Wang for
helpful discussions. We also would like to thank the anonymous
referee for his/her valuable suggestions on our work. This work is
supported in part by the National Natural Science Foundation of
China under Grants Nos. 90303004, 10533010, 10675136 and 10775180
and by the Chinese Academy of Science under Grant No.
KJCX3-SYW-N2. The work of RB is supported in part by an NSERC
Discovery Grant, by funds for the Canada Research Chairs Program,
and by an FQRNT Team Grant. RB thanks the KITPC for hospitality
during the period when a large part of the work on this project
was carried out.

\end{document}